\documentclass[twocolumn, prb, aps, superscriptaddress,longbibliography,showpacs,amsmath,amssymb,floatfix]{revtex4-2}                   
\usepackage{changes}

\usepackage{graphicx}
\usepackage{dcolumn}
\usepackage{bm}
\usepackage{graphicx}                                
\usepackage{amssymb}

\usepackage{amsmath}
\usepackage{epsfig}
\usepackage{xcolor}
\usepackage{tabu}
\usepackage{mathtools}
\usepackage[colorlinks,linkcolor=blue,anchorcolor=blue,citecolor=blue,urlcolor=blue]{hyperref}
\usepackage{physics}
\usepackage{float}
\usepackage{diagbox}
\usepackage{inputenc}

\begin{document}

\title{Hidden self-duality and exact mobility edges in quasiperiodic network models}

\author{Hai-Tao Hu}
\affiliation{Key Laboratory of Quantum Information, University of Science and Technology of China, Hefei 230026, China}
\affiliation{Anhui Province Key Laboratory of Quantum Network, University of Science and Technology of China, Hefei 230026, China}
\author{Xiaoshui Lin}
\affiliation{Key Laboratory of Quantum Information, University of Science and Technology of China, Hefei 230026, China}
\affiliation{Synergetic Innovation Center of Quantum Information and Quantum Physics, University of Science and Technology of China, Hefei 230026, China}
\author{Ai-Min Guo}
\affiliation{Hunan Key Laboratory for Super-microstructure and Ultrafast Process, School of Physics, Central South University, Changsha 410083, China}
\author{Guangcan Guo}
\affiliation{Key Laboratory of Quantum Information, University of Science and Technology of China, Hefei 230026, China}
\affiliation{Anhui Province Key Laboratory of Quantum Network, University of Science and Technology of China, Hefei 230026, China}
\affiliation{Hefei National Laboratory, University of Science and Technology of China, Hefei 230088, China}
\affiliation{Synergetic Innovation Center of Quantum Information and Quantum Physics, University of Science and Technology of China, Hefei 230026, China}
\author{Zijing Lin}
\email{zjlin@ustc.edu.cn}
\affiliation{Hefei National Laboratory, University of Science and Technology of China, Hefei 230088, China}
\affiliation{Department of Physics, University of Science and Technology of China, Hefei, Anhui 230026, China}
\author{Ming Gong}
\email{gongm@ustc.edu.cn}
\affiliation{Key Laboratory of Quantum Information, University of Science and Technology of China, Hefei 230026, China}
\affiliation{Anhui Province Key Laboratory of Quantum Network, University of Science and Technology of China, Hefei 230026, China}
\affiliation{Hefei National Laboratory, University of Science and Technology of China, Hefei 230088, China}
\affiliation{Synergetic Innovation Center of Quantum Information and Quantum Physics, University of Science and Technology of China, Hefei 230026, China}

\date{\today}

\begin{abstract}
In one-dimensional quasiperiodic systems, only a few models with exact mobility edges (MEs) have been constructed using generalized self-duality theory, Avila's global theory, or the renormalization group method. This raises an intriguing question that whether we can realize more physical models with exact solvable MEs. In this work, we uncover the hidden self-duality within a class of quasiperiodic network models constituted by periodic and quasiperiodic sites. Although the original Hamiltonians appear to lack self-duality, their effective Hamiltonians obtained by integrating out the periodic sites exhibit self-duality, which yield MEs. The well-studied mosaic model, which is the simplest case of quasiperiodic network models, was previously thought to exhibit MEs due to the absence of self-duality, but we show that they actually arise from the hidden self-duality. Using the effective Hamiltonian, we further introduce the concept of resonant states to understand the shape of MEs. Finally, we present in detail how to determine the MEs in various network models, including some non-Hermitian models, based on the hidden self-duality. These predictions can be experimentally realized using optical and acoustic waveguide arrays. Our work can greatly advance our understanding of MEs in Anderson transition.
\end{abstract}

\maketitle

{\em Introduction.---}The mobility edge (ME) in disordered models, since the discovery of Anderson localization \cite{Anderson1958Diffusion, Thouless1974Electrons, Mott1987mobility, Schwartz2007Transport, Roati2007Anderson, Billy2008Anderson}, has became one of the most fundamental concepts in condensed matter physics. In these models, the wave functions, induced by interference during scattering by the random potentials, can exhibit significantly different behaviors at some particular energy $E$, below or above which the wave functions are either extended or localized, respectively. In three-dimensional disordered models, Anderson transition can occur at finite disorder strength, naturally yielding MEs \cite{evers2008anderson}. However, in one- and two- dimensional models with  random potentials,  almost all states are localized, and MEs are absent. These results are well described by the scaling theory \cite{abrahams1979scaling,MacKinnon1981OneParameter,Sarker1981Scaling}.

\begin{figure}
\includegraphics[width=0.48\textwidth]{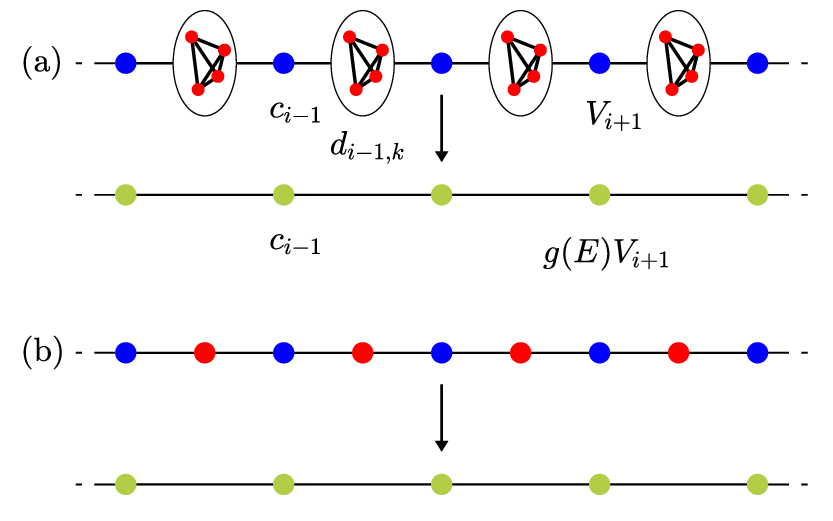}
\caption{(a) The generalized one-dimensional quasiperiodic network model constituted by periodic network sites between quasiperiodic sites and its reduced effective quasiperiodic model. Here, $V_{i+1}$ and $g(E) V_{i+1}$ denote the potentials before and after integrating out the periodic network sites. (b) The quasiperiodic mosaic model and its reduced effective quasiperiodic model. In both figures, the blue, red, and green solid points denote the lattice sites with quasiperiodic, periodic, and effective energy-dependent potentials, respectively.}
\label{fig-fig1} 
\end{figure}

\begin{figure*}
\includegraphics[width=0.95\textwidth]{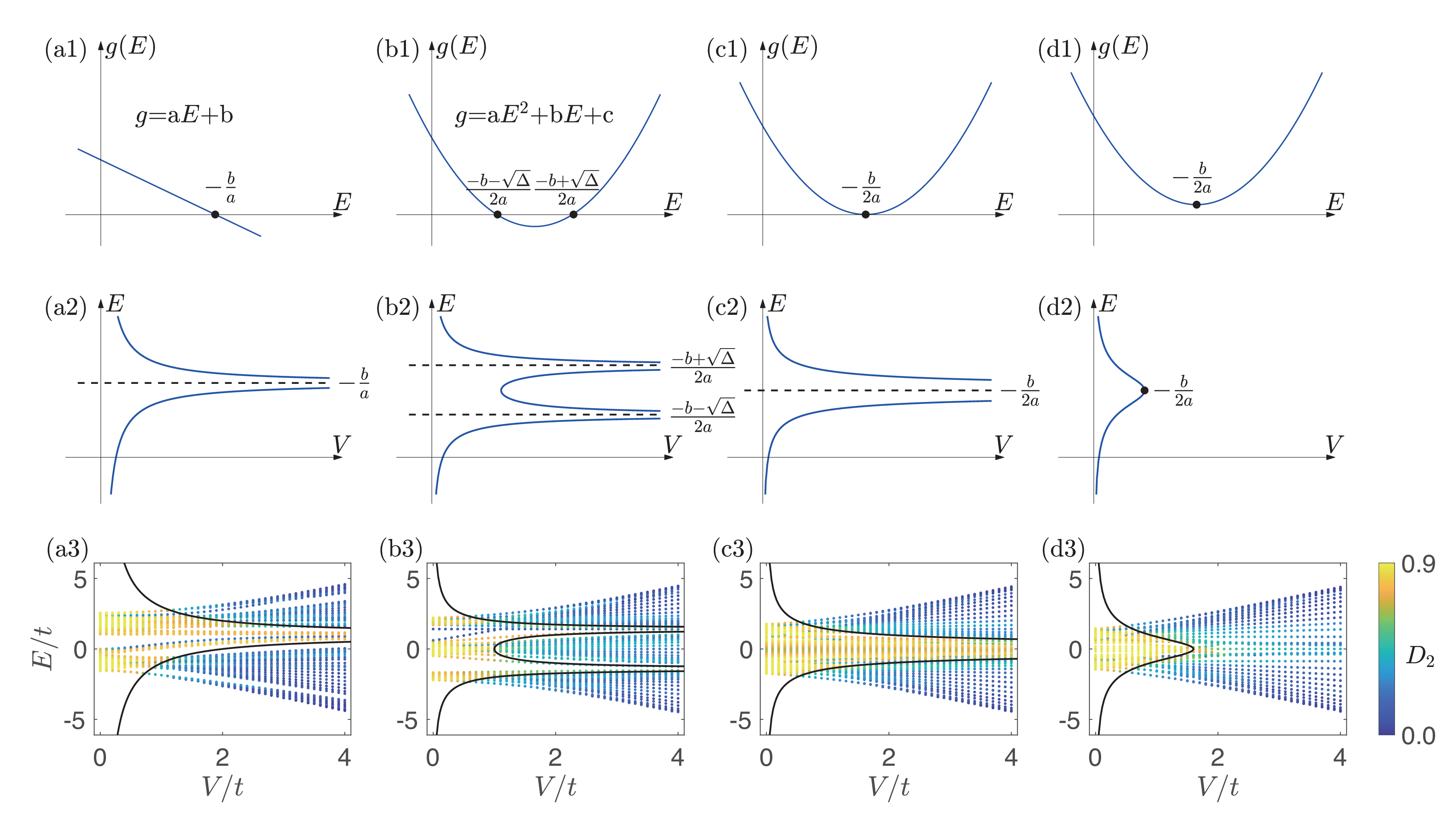}
\caption{ Energy-dependent $g(E)$ and their corresponding MEs. (a1)-(d1) illustrate linear and quadratic $g(E)$ and (a2)-(d2) show their possible MEs. The shape of the MEs depends on the root structure of $g(E) = 0$. Realizations of MEs in the quasiperiodic mosaic model are presented in (a3) with $\kappa = 2$, $U_1 = t$, (b3) with $\kappa = 3$, $U_1 = -U_2 = t$, (c3) with $U_1 = -U_2 = it$, and (d3) with $U_1 = -U_2 = 1.5it$. Non-Hermitian models in (c3) and (d3) exhibit parity-time (PT) symmetry.}
\label{fig-fig2}
\end{figure*}

Nevertheless, Anderson  transitions and the associated MEs can be realized in low-dimensional models with quasiperiodic potentials \cite{aubry1980analyticity, biddle2010mobility, Harper1955Single, Liu2024Dissipation, Chang2025Observation, hu2025divergent}, in which  the most widely studied one is the Aubry-Andr\'{e}-Harper (AAH) model \cite{aubry1980analyticity, Harper1955Single}  realized in a two-dimensional square lattice with perpendicular magnetic field. In this way, the localization-delocalization transition is intimately related to the Hofstadter butterfly and its topological edge states \cite{Hofstadter1976Energy}. In the AAH model, the Hamiltonian contains a self-dual point without ME, that is, there is a critical parameter $V_c$, for which when $|V| < V_c$ all states are extended and for $|V| > V_c$ all states are localized. At the critical point $|V| = V_c$, all states are critical, exhibiting multifractal wave functions and discrete eigenvalues (or point spectra) \cite{wilkinson1984critical, Siebesma1987Multifractal, Abe1987Fractal}. Many unique properties about their MEs have been discovered in various generalized AAH models \cite{kraus2012topological, cai2013topological, ganeshan2013topological, Wang2016Phase, Luschen2018Mobility, loughi2019topological, Loughi2019phase, Goblot2020criticality, Duthie2021theory, Roy2021Reentrant, Zhai2021Cascade, Duncan2024Critical, Borgnia2022Rational, Borgnia2023Localization, Longhi2024Dephasing}. In the presence of hopping beyond neighboring sites \cite{Gopalakrishnan2017quasiperiodic, deng2019quasicrystals, Fraxanet2022Localization, Xia2022mobility, Wang2023Fate}, or replacing the AAH potential using some other quasiperiodic potentials \cite{Soukoulis1982Localization, Griniasty1988Localization, Li2017Mobility, Yao2019Critical, Das1988Mobility, Thouless1988Localization, Das1990Localization, Liu2018Mobility, hu2025exact}, or introducing a mixture of clean and disordered chains \cite{Rossignolo2019Localization, Lin2023critical, Lin2024Fate}, the MEs can be realized, in which the systems display both conducting and insulating phases depending on the energy $E$ \cite{sil2008transition}.  Although exact MEs in some specific quasiperiodic models have been obtained through duality transformations \cite{aubry1980analyticity, biddle2010mobility}, Avila’s global theory \cite{Avila2015Global, Wang2020Quasiperiodic}, or renormalization groups method \cite{Goncalves2023Critical, Goncalves2023Renormalization}, the underlying mechanism for MEs remains unclear due to the involved complicated mathematical techniques. 

In this manuscript, we reveal the hidden self-duality within a class of quasiperiodic network models composed of periodic and quasiperiodic sites, with the simplest realization being the quasiperiodic mosaic models \cite{Wang2020Quasiperiodic, Liu2021Exact2, wang2022topological, zhou2023mobility, dai2023multifractality}. Previously, the MEs in the quasiperiodic mosaic model were attributed to the breaking of self-duality \cite{Wang2020Quasiperiodic}. However, we find that its effective energy-dependent Hamiltonian, obtained by integrating out the periodic sites, actually possesses a hidden self-duality. Next, we generalized this idea to more complicated quasiperiodic network models, showing that the behavior of MEs can be characterized by two rational polynomials $f(E)$ and $g(E)$, which represents the effective energy and effective potential, respectively. Based on the effective Hamiltonian, we can introduce the concept of resonant state from the root of $g(E) = 0$ to explain the shape of MEs. Finally, we determine the exact MEs in various quasiperiodic network models through hidden self-duality. The major merit of this idea is that the MEs can be constructed in a much easier way by engineering of these two functions without complicated mathematical tricks \cite{Wang2020Quasiperiodic, Goncalves2023Critical}. The quasiperiodic models constructed in this work can be realized using the optical and acoustic waveguide arrays in experiments \cite{Wang2022EdgeState, Xiao2021Observation, Wang2022Thouless, Wiersma1997Localization, Wiersma2013Disordered, Kondakci2015photonic, Wang2020Localization, Yu2021Engineered}, which can greatly advance our understanding of the origin of MEs. 

\begin{table*}[!htbp]
  \centering
  \caption{The structure of three quasiperiodic network models and their corresponding $g(E)$ and $f(E)$ for the effective Hamiltonian. Numerical verification of these MEs for three possible $V_i$ in Eq.~(\ref{eq-threepossibleVj}) are presented in Fig.~\ref{fig-fig3} and Supplemental Material \cite{supplematHuHaiTao}. The resonance states, independent of $V_i$, are given by $g(E) = 0$.}
  \begin{tabular}{ |c|  c|  c|  c|  }
    \hline \hline
    Structure &
    \begin{minipage}[b]{0.6\columnwidth}
		\centering
		\raisebox{-.5\height}{\includegraphics[width=0.8\linewidth]{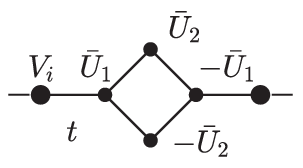}}
	\end{minipage} 
    & \begin{minipage}[b]{0.5\columnwidth}
		\centering
		\raisebox{-.5\height}{\includegraphics[width=0.7\linewidth]{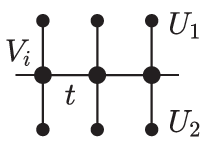}}
	\end{minipage} 
    & \begin{minipage}[b]{0.5\columnwidth}
		\centering
		\raisebox{-.5\height}{\includegraphics[width=0.95\linewidth]{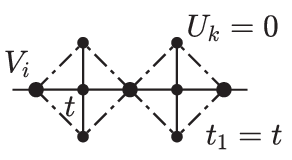}}
	\end{minipage}
    \\ \hline
    
    \rule[0mm]{0pt}{6mm}
     
    $g(E)$ & $\dfrac{E^4 - E^2(\bar{U}_1^2+\bar{U}_2^2+4t^2) + \bar{U}_1^2 \bar{U}_2^2}{2E}$ & $1$ & $\dfrac{E^2-2t^2}{3E-4t}$
    \\ 
    \rule[-3mm]{0pt}{9mm}
     
    $f(E)$ & $\dfrac{E^5 - E^3(\bar{U}_1^2+\bar{U}_2^2+6t^2) + E(2\bar{U}_2^2t^2+\bar{U}_1^2\bar{U}_2^2+4t^4)}{2E}$ & $E-\dfrac{t^2}{E-U_1}-\dfrac{t^2}{E-U_2}$ & $\dfrac{E^3-8Et^2+8t^3}{3E-4t}$ 
    \\ \hline \hline
  \end{tabular}
\label{table}
\end{table*}

{\em From self-duality to hidden self-duality.---}We begin with the following general \textcolor{black}{single-particle} Hamiltonian 
\begin{equation}
H = H_{\rm q} + H_{\rm p} + H_{\rm c}, 
\end{equation}
by partitioning the entire system into periodic subsystem with Hamiltonian $H_{\rm p}$,  quasiperiodic subsystem with Hamiltonian $H_{\rm q}$, and their couplings $H_{\rm c}$. In this way, the above Hamiltonian can be rewritten in the following form 
\begin{align}
\begin{pmatrix}
H_{\rm q} & H_{\rm c} \\
H_{\rm c}^\dagger & H_{\rm p}
\end{pmatrix}
\begin{pmatrix}
\psi_{\rm q} \\
\psi_{\rm p}
\end{pmatrix}
=
E
\begin{pmatrix}
\psi_{\rm q} \\
\psi_{\rm p}
\end{pmatrix},
\end{align}
where $\psi_\text{p}$ and $\psi_\text{q}$ represent the wave functions of the periodic and quasiperiodic sites, respectively, and $E$ is the eigenvalue. Integrating out the periodic wave functions yields the effective Hamiltonian for the quasiperiodic sites as $E\psi_{\rm q} = H_{\rm eff}' \psi_{\rm q}$,  where $H_{\rm eff}' = H_\text{q} + H_\text{c} {1\over E - H_\text{p}} H_\text{c}^\dagger$. Defining $1/(E-H_{\rm p}) = C/\text{det}(E-H_{\rm p})$, where $C$ is the adjoint of the matrix $(E-H_{\rm p})$ based on co-factors, we can obtain an effective Hamiltonian 
\begin{equation}
H_\text{eff} \psi_q = f(E)\psi_q,
\end{equation}
for the quasiperiodic sites. A possible choice is $H_\text{eff} = g(E) H_{\rm q} + H_{\rm c}^\dagger C H_{\rm c} + \mathcal{E}$, where $f(E) = \text{det}(E-H_{\rm p}) E + \mathcal{E}$, $g(E) = \text{det}(E-H_{\rm p})$ and $\mathcal{E}$ may depend on energy $E$. Examples of the effective Hamiltonian calculation can be found in the Supplemental Material \cite{supplematHuHaiTao}.

For the model presented in Fig.~\ref{fig-fig1}(a) where all quasiperiodic sites are disconnected, we can choose the quasiperiodic potential as the AAH model, given by $H_{\rm q} = \sum_i V \cos(2 \pi \alpha i) c_i^\dagger c_i$, where $c_{i}^{\dagger}$ $(c_{i})$ is the creation (annihilation) operator at site $i$, $V$ is the strength of quasiperiodic potential, and $\alpha$ is an irrational number. Consequently, we can obtain an effective Hamiltonian with energy-dependent parameters
\begin{eqnarray}
H_{\rm eff} = \sum_i g(E) V \cos(2 \pi \alpha i) c_i^\dagger c_i + t^\kappa (c_{i}^\dagger c_{i+1} + {\rm H.c.}).
\label{eq-model1}
\end{eqnarray}
Here, $t$ describes the nearest-neighbor hopping strength, $g(E)$ is the effective potential strength, $f(E)$ is the effective eigenvalue of $H_\text{eff}$, and $\kappa \ge 2$. $f(E)$ and $g(E)$ are determined by the structure of the network sites. After applying a Fourier transformation to momentum space $b_k \sim \sum_i {\rm exp}(i 2 \pi \alpha k i) c_i$ \cite{aubry1980analyticity, Wang2021Duality, Goncalves2022Hidden}, the Hamiltonian becomes $H_\text{eff}^k = \sum_k {g(E)V}/{2} (b_{k+1}^\dagger b_k + {\rm H.c.}) + 2t^\kappa \cos(2 \pi \alpha k) b_k^\dagger b_k$. Through self-duality, we have the following theorem.

{\bf Theorem}: Any quasiperiodic network model that can be mapped to the effective Hamiltonian in Eq.~(\ref{eq-model1}) exhibits MEs at \cite{aubry1980analyticity, Liu2020Generalized, Wang2020Quasiperiodic}
\begin{equation}
g(E) V = \pm 2t^\kappa.
\label{eq-theorem}
\end{equation}
By engineering the function $g(E)$, arbitrary forms of MEs can be realized immediately.

{\em Resonant state and shape of the mobility edges.---}The above theorem forms a central idea to be presented in this work. From the effective Hamiltonian, it is evident that the states satisfying $g(E) = 0$ are fully extended. We refer to them as the resonant state, since it tends to pass through the quasiperiodic chain regardless of $V_i$. For the resonant state, these incommensurate sites act as node of the wave functions. Furthermore, the shape of the MEs is almost determined by the number of resonant states $N_{\rm r}$, which equals to the number of real roots in $g(E) = 0$. Thus the number of possible MEs is at most $N_{\rm r}+1$. In the case where $g(E) \equiv 1$, we recover the well-known results of the AAH model, without resonant state and ME, corresponding to $N_{\rm r} = 0$. 

To illustrate the underlying physics, consider several concrete polynomials. For $g(E) = aE + b$, there exists a special state with $E = -b/a$ from $g(E) = 0$, 
making the chain independent of the quasiperiodic potential. In this scenario, only two MEs exist, since all states are extended when $V = 0$, but only the resonant state remains extended when $V$ is infinite [see Fig.~\ref{fig-fig2}(a2)]. This result can also be derived from the self-duality condition in Eq.~(\ref{eq-theorem}). For $g(E) = aE^2 + bE + c$, the scenario depends on the discriminant $\Delta = b^2-4ac$. (I) When $\Delta > 0$, there are two resonant states at $ E = (-b \pm \sqrt{\Delta})/(2a)$, resulting in three disconnected MEs across the entire $V$ region in Fig.~\ref{fig-fig2}(b2); (II) When $\Delta = 0$, there is only one resonant state at $E = -b/2a$, leading to two MEs across the entire $V$ as shown in Fig.~\ref{fig-fig2}(c2); (III) When $\Delta < 0$, no resonant state is allowed, yielding one ME in Fig.~\ref{fig-fig2}(d2). In this way, the root structure of the $g(E) = 0$ completely determine the structure of the MEs. Above, we discuss these physics based on the effective Hamiltonians, and in following, we will focus on its application in concrete physical models and illustrate their MEs through hidden self-duality. 

{\em Hidden self-duality in quasiperiodic mosaic model.---}In the previous literature, the MEs in mosaic models were attributed to the breaking of self-duality \cite{Wang2020Quasiperiodic}. However, one of the key findings in this manuscript is that the MEs observed in these models actually come from the hidden self-duality. To illustrate this, we consider
\begin{align}
\mathcal{H} = &-t \sum_{i} (c_{i}^\dagger d_{i,1} + \sum_{k=1}^{\kappa-2} d_{i,k}^\dagger d_{i,k+1} + d_{i,\kappa-1}^\dagger c_{i+1} + \mathrm{H.c.} ) \nonumber \\
&+ \sum_{i} (V_i c_i^\dagger c_i + \sum_{k=1}^{\kappa-1} U_{k} d_{i,k}^\dagger d_{i,k}),
\label{eq1}
\end{align}
with $V_i = V\cos(2\pi\alpha i)$.
Here, $\kappa \ge 2$, the sites without incommensurate potential can be seen as the simplest realization of periodic network, and $V$ and $U_k$ denote quasiperiodic and periodic potential strengths, respectively. To confirm the hidden self-dual points, we numerically determine the MEs from the inverse participation ratio (IPR) defined as ${\rm IPR^{n}} = \sum_{i = 1}^{L} |\psi_{i}^{n}|^4$ \cite{evers2008anderson}, where $\psi_i^n$ denotes the $n$-th normalized wave function amplitude at site $i$. Additionally, the associated fractal dimension is given by $D_2^{n} = - \lim _{L \rightarrow \infty} [{\rm ln}({\rm IPR^{n}}) / {\rm ln}L]$ \cite{Lin2023critical, Lin2024Fate}. $D_2$ tends to 1 for extended states and 0 for localized states, while $0 < D_2 < 1$ characterizes multifractal states. Without loss of generality, hereafter we set $\alpha = (\sqrt{5} - 1) / 2$ in all our numerical simulations.

Let us first consider the case with $\kappa = 2$, in which we have $- t (d_{i-1}  + d_{i}) + V_{i} c_{i} = E c_{i}$, for all $i \in \mathbb{Z}$. This yields the effective equation for the quasiperiodic sites as  
\begin{equation}
g(E) V \cos(2 \pi \alpha i) c_i + t^2(c_{i-1} + c_{i+1}) = f(E) c_i,
\label{eq-motionkapp2}
\end{equation}
with polynomials $f(E) = E^2 -2t^2 -E U_1$, and $g(E) = E-U_1$. This process is illustrated in Fig.~\ref{fig-fig1}(b). A key observation is that integrating out the periodic network sites results in an energy-dependent potential $g(E) V_i$ in the effective model. When $E = U_1$, the effective equation becomes independent of $V_i$, indicating the corresponding state to be the resonant state, where the wave functions occupy only the periodic sites, hence $c_i = 0$ [Fig.~\ref{fig-fig2}(a3)]. Note that here $\sum_i |c_i|^2$ is not necessary to be normalized. According to Eq.~(\ref{eq-theorem}), the self-dual points happen at $V(E -U_1) = \pm 2t^2$, naturally yielding energy dependent MEs. 

The same approach can be applied to $\kappa = 3$. In this condition, we have the same equation as Eq.~(\ref{eq-motionkapp2}), with $f(E) = -E^3 + (U_1+U_2)E^2 + (3t^2-U_1U_2)E-t^2(U_1+U_2)$, and $g(E) = -E^2 + (U_1 + U_2) E - U_1 U_2 + t^2$, yielding MEs at $g(E) V = \pm 2t^3$. Obviously when $U_1$ and $U_2$ are both real, there must be two unequal real solutions for $g(E) = 0$, indicating of two resonant states at $2E = U_1 + U_2 \pm \sqrt{4t^2 + (U_1 - U_2)^2}$, which ensures that there are three MEs in the system [Fig.~\ref{fig-fig2}(b3)]. If we consider complex $U_1$ and $U_2$ with parity-time (PT) symmetry \cite{Bender1998Real}, the shape of the MEs depends on the discriminant $\Delta = (U_1 - U_2)^2 + 4 t^2$. When $U_1 = -U_2 = it$ with $\Delta = 0$, two MEs can be observed [see Fig.~\ref{fig-fig2}(c3)], whereas when $U_1 = -U_2 = 1.5it$ with $\Delta < 0$, only one ME can be realized [see Fig.~\ref{fig-fig2}(d3)].

\begin{figure}
\includegraphics[width = 0.45\textwidth]{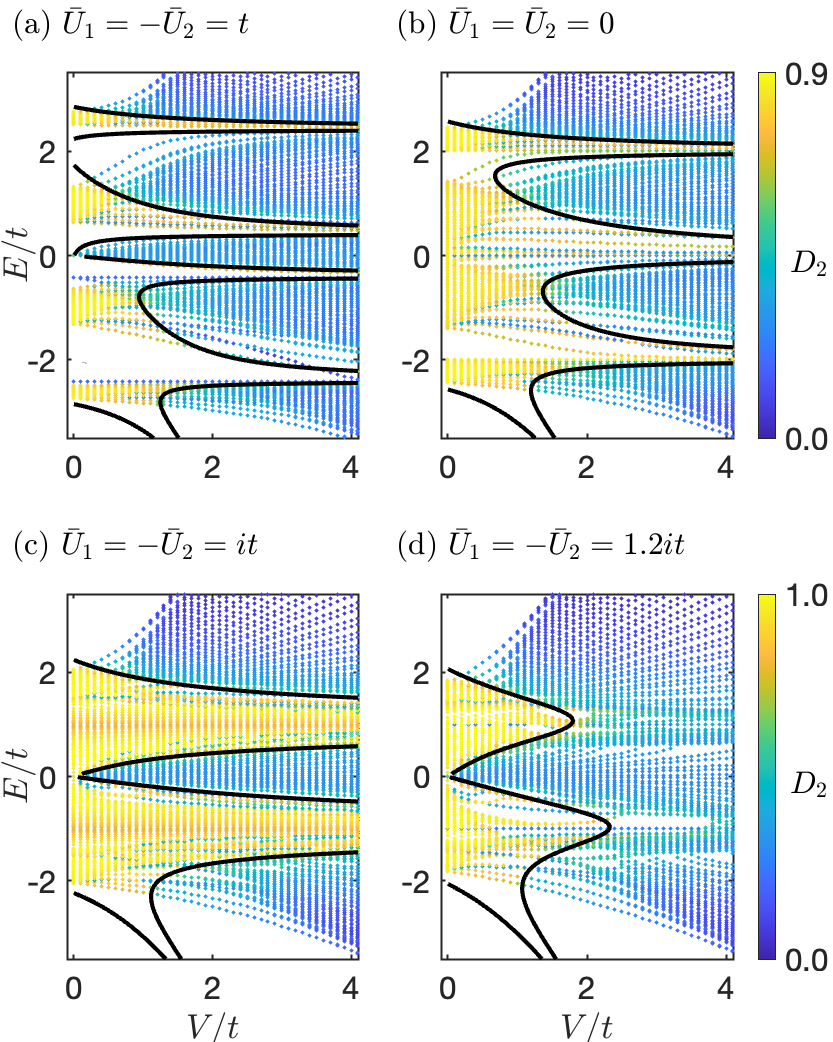}
\caption{
Fractal dimension $D_2$ of different eigenstates versus $E$ and potential strength $V$ for (a) $\bar{U}_1 = -\bar{U}_2 = t$, (b) $\bar{U}_1 = \bar{U}_2 = 0$, (c) $\bar{U}_1 = -\bar{U}_2 = it$, and (d) $\bar{U}_1 = -\bar{U}_2 = 1.2it$ using the quasiperiodic potential $V_i = V \cos (2 \pi \alpha i)/[1-b \cos (2 \pi \alpha i)]$. The periodic network is given by the first structure of Table \ref{table}. The solid lines denote the MEs $b f(E) = \pm 2t - V g(E)$ with $f(E)$ and $g(E)$ given in Table \ref{table}. The other parameters are $\alpha = (\sqrt{5} - 1) / 2$, $b = 0.5$ and $L = 600$.}
\label{fig-fig3}
\end{figure}

{\em Hidden self-duality in quasiperiodic network model.---}The above idea of deriving MEs via hidden self-duality can be generalized to more quasiperiodic network models. In Table \ref{table}, we 
summarize three representative network structures along with their corresponding $g(E)$ and $f(E)$. In these models, $V_i$ can take various forms of quasiperiodic potentials, corresponding to different hidden self-duality points \cite{Ganeshan2015Mobility, Bender1998Real, Liu2020Generalized}. 

To illustrate this idea, in the main text we consider a quasiperiodic diamond model (the first network structure in Table \ref{table}) with the Hamiltonian as 
\begin{align}
\mathcal{H} = & -t \sum_{i} (c_{i}^\dagger d_{i,1} + d_{i,3}^\dagger c_{i+1} + \sum_{k = 1}^{4} d_{i,k}^\dagger d_{i,k+1} + \mathrm{H.c.} ) \nonumber \\
& + \sum_{i} (V_i c_i^\dagger c_i + \sum_{k = 1}^4 U_k d_{i,k}^\dagger d_{i,k}),
\label{eq1}
\end{align}
where $d_{i,k}$ is the annihilation operator at the $k$-th site of the $i$-th periodic network, $U_k$ denotes the on-site energy, and $V_i$ represents the quasiperiodic potential strength. For simplicity, we choose $U_1 = -U_3 = \bar{U}_1$, $U_2 = -U_4 = \bar{U}_2$. For three possible forms of quasiperiodic potential 
\begin{equation}
V_i = V {\rm cos}(\theta_i), \quad {V \cos(\theta_i) \over 1 - b \cos(\theta_i)}, \quad {V \over 1 - b \exp(i\theta_i)},
\label{eq-threepossibleVj}
\end{equation}
with $\theta_i = 2\pi \alpha i$, we can determine the corresponding MEs at 
\begin{equation}
g(E) V = \pm 2t^4, \quad bf(E) = \pm 2t^4 - V g(E), \quad f(E) = b + {1 \over b},
\label{eq-MEs}
\end{equation}
where the expressions of $g(E)$ and $f(E)$ are determined by the structures in Table \ref{table}. The rigorous derivation and numerical verification of Eq.~(\ref{eq-MEs}) are presented in the Supplemental Material \cite{supplematHuHaiTao}. In Fig.~\ref{fig-fig3}, we plot the fractal dimension of the corresponding eigenstates as a function of energy $E$ and potential strength $V$ for various $\bar{V}_1$ and $\bar{V}_2$. The solid lines in the figures represent the exact MEs derived from hidden self-duality, which agree well with the numerical results. It is evident that the shape of the MEs highly depends on the roots of $g(E) = 0$, which can be controlled and engineered by the parameters involved in the Hamiltonians and the structure of the networks. The numerical verifications of the results for the second and third network models are presented in Supplemental Material \cite{supplematHuHaiTao}. 

{\em Conclusion.---}This work reveals the hidden self-duality in a class of quasiperiodic network models and demonstrate their corresponding resonant states by constructing their effective Hamiltonians. The key idea is that after integrating out the periodic sites, the effective quasiperiodic models contain energy-dependent system parameters, which naturally yield energy-dependent MEs. As a result, while the original models lack self-duality, their effective models can possess self-duality --- termed as hidden self-duality --- for MEs  \cite{Wang2020Quasiperiodic, zhou2023mobility}. This work opens a totally new avenue for engineering of Anderson transitions and their related MEs in quasiperiodic network models with hidden self-duality, which can be realized and verified using optical and acoustic waveguide arrays \cite{Wang2022EdgeState, Zeng2020Topological, Chen2021LandauZener, Wiersma1997Localization, Wiersma2013Disordered, Kondakci2015photonic, Wang2020Localization, Yu2021Engineered}. In these experiments, the network sites can be easily realized by designing their structures and the couplings between the waveguides. We hope the network models presented in this work can provide new insight into the fundamental origin of MEs during Anderson transition \cite{Zhang2022Lyapunov, Wang2023mobility, Padhan2024Complete}. 
 \textcolor{black}{Finally, with some proper modifications, the similar idea of constructing an effective Hamiltonian with energy-dependent parameters by reducing the Hilbert space dimension may also be applied to study the quasiperiodic many-body systems for intriguing physics, including many-body thermalization \cite{Rahul2015Many, Abanin2019Colloquium}, entanglement \cite{ Islam2015Measuring, Lukin2019Probing}, and criticality \cite{Vosk2015Theory,Li2015Localization, Modak2015Many, Fischer2016Dynamics, Choi2016Exploring}.}

We thank the insightful discussion with Prof. Zhenyu Zhang, Prof. Qian Niu, Prof. Qi Zhou, and Xin-Chi Zhou. This work is supported by the Strategic Priority Research Program of the Chinese Academy of Sciences (Grant No. XDB0500000), the National Natural Science Foundation of China (Grant No. 12374017, No. 12074362 and No. U23A2074) and the Innovation Program for Quantum Science and Technology (2021ZD0303303, 2021ZD0301200, 2021ZD0301500). A.-M. G. is supported by the NSFC (Grant Nos. 12274466 and 11874428) and the Hunan Provincial Science Fund for Distinguished Young Scholars (Grant No. 2023JJ10058).

{\em Note added.---}After this work was submitted, several related
preprints \cite{li2025multifractal, zhang2025emergent, wang2025exact, lin2025topological} appeared, which use the same method to obtain the effective Hamiltonians and characterize their phase transitions.

\setcounter{equation}{0} \setcounter{figure}{0}
\setcounter{table}{0} 
\renewcommand{\theparagraph}{\bf}
\renewcommand{\thefigure}{S\arabic{figure}}
\renewcommand{\theequation}{S\arabic{equation}}

\onecolumngrid
\flushbottom
\newpage

\begin{center}
{\Large\bfseries Supplementary Materials for ``Hidden self-duality and exact mobility edges in quasiperiodic network models''\par}
\vspace{0.5em}
\end{center}

This Supplemental Material provides additional details for the results presented in the main text. In Sec.~\ref{sec-ApendixA}, we summarize the model that can be reduced to an effective nearest-neighbor hopping model. Section \ref{sec-ApendixB} presents the details of deriving the effective Hamiltonian. In Sec.~\ref{sec-ApendixC}, we derive the generalized self-dual points for two quasiperiodic models and present the corresponding numerical results. In Sec.~\ref{sec-ApendixD}, we outline the complete procedure for obtaining the mobility edges (MEs) through hidden self-duality. In Sec.~\ref{sec-ApendixE}, we verify the MEs for additional structures in the main text. In Sec.~\ref{sec-ApendixF}, we investigate the effect of the phase term in the quasiperiodic potential on localization properties. Finally, we perform a finite-size scaling analysis of the mean fractal dimension in Sec.~\ref{sec-ApendixG}, demonstrating that our results are exact at the thermodynamic limit. 

\section{Two kinds of quasiperiodic network models}
\label{sec-ApendixA}
In this work, the major idea is to divide the whole system into periodic and quasiperiodic sites. By integrating out the periodic sites, we can obtain an effective Hamiltonian with energy dependent quasiperidic potentials. The key insight of our approach is to treat the effective model as a real physical system and regard the energy $E$ as a tunable parameter. In specific cases, we identify some special properties of the effective Hamiltonian, such as self-duality, which enables the exact determination of its MEs.

\begin{figure}
\centering\includegraphics[width=0.6\textwidth]{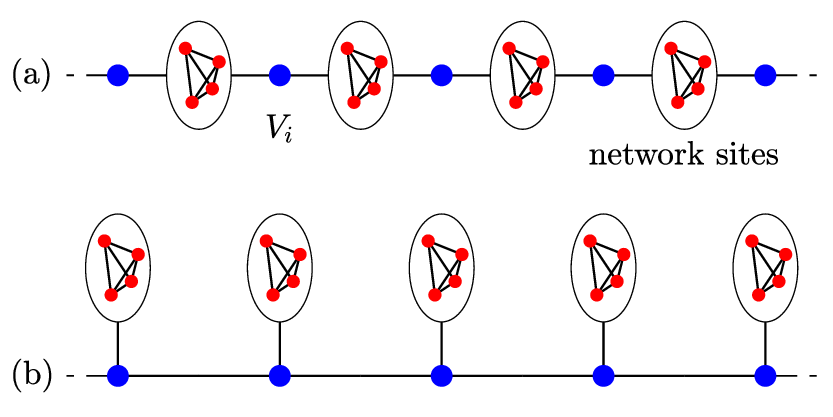}
\caption{Two kinds of quasiperiodic network models, which can be mapped to an effective nearest-neighbor chain.}

\label{fig-figsup1}
\end{figure}

In the main text, we present several models where the corresponding effective Hamiltonians involve only nearest-neighbor hopping. These models can be broadly classified into two categories (see Fig.~\ref{fig-figsup1}), referred to as network models. For the configuration in Fig.~\ref{fig-figsup1}(a), the Hamiltonian can be expressed as
\begin{align}
H = 
\begin{pmatrix}
\ddots & & & & & & & \\
 & V_{i-1} & H_{\rm c1} & & & & & \\
 & H_{\rm c1}^\dagger & H_{\rm p} & H_{\rm c2} & & & & \\
 & & H_{\rm c2}^\dagger & V_i & H_{\rm c1} & & & \\
 & & & H_{\rm c1}^\dagger & H_{\rm p} & H_{\rm c2} & & \\
 & & & & H_{\rm c2}^\dagger & V_{i+1} & H_{\rm c1} & \\
 & & & & & H_{\rm c1}^\dagger & H_{\rm p} & \\
 & & & & & & & \ddots
\end{pmatrix}
,
\end{align}
where $V_i$ is the quasiperiodic potential at site $i$, $H_{\rm p}$ describes the Hamiltonian of the periodic sites, and $H_{\rm c1}$ and $H_{\rm c2}$ denote the couplings between the quasiperiodic and periodic sites. Here, $V_i$ is a scalar, while $H_{\rm p}$, $H_{\rm c1}$ and $H_{\rm c2}$ are matrices, which can take arbitrary forms
\begin{align}
H_{\rm p} = 
\begin{pmatrix}
U_1 & t_{1,2} & \cdots & t_{1,\kappa-2} & t_{1,\kappa-1} \\ 
t_{1,2}^\dagger & U_2 & \cdots & t_{2,\kappa-2} & t_{2,\kappa-1} \\
\vdots & \vdots & \ddots & \vdots & \vdots \\
t_{1,\kappa-2}^\dagger & t_{2,\kappa-2}^\dagger & \cdots & U_{\kappa-2} & t_{\kappa-2,\kappa-1} \\
t_{1,\kappa-1}^\dagger & t_{2,\kappa-1}^\dagger & \cdots & t_{\kappa-2,\kappa-1}^\dagger & U_{\kappa-1}
\end{pmatrix}
, \quad
H_{\rm c1/c2} = 
\begin{pmatrix}
t_1 & t_{2} & \cdots & t_{\kappa-2} & t_{\kappa-1}
\end{pmatrix}
.
\end{align}

Similarly, the Hamiltonian for the second configuration in Fig.~\ref{fig-figsup1}(b) is given by
\begin{align}
H =
\begin{pmatrix}
\ddots & & & & & & & \\
 & V_1 & H_{\rm c} & t & & & & \\
 & H_{\rm c}^\dagger & H_{\rm p} & & & & & \\
 & t & & V_2 & H_{\rm c} & t & & \\
 & & & H_{\rm c}^\dagger & H_{\rm p} & & & \\
 & & & t & & V_3 & H_{\rm c} & \\
 & & & & & H_{\rm c}^\dagger & H_{\rm p} & \\
 & & & & & & & \ddots
\end{pmatrix}
,
\end{align}
where $t$ serves as the energy unit.

For both types of models, their Hamiltonian can be written as $H_\text{eff} \psi = f(E) \psi$ with effective nearest-neighbor hopping Hamiltonian (see Sec.~\ref{sec-ApendixB} for details)
\begin{align}
H_{\rm eff} = \sum_i g(E) V_i c_i^\dagger c_i + t^\kappa (c_{i}^\dagger c_{i+1} + {\rm H.c.}).
\label{eq-effHami}
\end{align}
Here, $V_i$ is an quasiperiodic potential of any form. The forms of $f(E)$ and $g(E)$ for various models can be found in the main text. Thus the energy $E$ acts as a control parameter governing both potential modulation and the effective eigenenergy $f(E)$. The self-duality in Eq.~(\ref{eq-effHami}) enables the determination of exact ME as detailed in Sec. \ref{sec-ApendixD}.

\section{Details on obtaining the effective model}
\label{sec-ApendixB}
In this section, we present the details on how to integrate out the periodic sites and obtain an effective nearest-neighbor hopping Hamiltonian [see Eq.~(\ref{eq-effHami})].
Let us consider a finite quasiperiodic mosaic model ($L = 6$) with periodic boundary condition, whose Hamiltonian reads
\begin{align}
H =
\begin{pmatrix}
V_1 & 0 & 0 & t & 0 & t \\ 
0 & V_2 & 0 & t & t & 0 \\
0 & 0 & V_3 & 0 & t & t \\
t & t & 0 & U_1 & 0 & 0 \\
0 & t & t & 0 & U_1 & 0 \\
t & 0 & t & 0 & 0 & U_1
\end{pmatrix}
\label{eq-hami1},
\end{align}
acting on the eigenstate $(\psi_{{\rm q},1}, \psi_{{\rm q},2}, \psi_{{\rm q},3}, \psi_{{\rm p},1}, \psi_{{\rm p},2}, \psi_{{\rm p},3})^{\rm T}$. It can be rewritten in block form as
\begin{align}
H =
\begin{pmatrix}
H_{\rm q} & H_{\rm c} \\ 
H_{\rm c}^\dagger & H_{\rm p}
\end{pmatrix}
,
\end{align}
where the submatrices are defined by
\begin{align}
H_{\rm q} =
\begin{pmatrix}
V_1 & 0 & 0 \\ 
0 & V_2 & 0 \\
0 & 0 & V_3
\end{pmatrix}
,
H_{\rm p} =
\begin{pmatrix}
U_1 & 0 & 0 \\ 
0 & U_1 & 0 \\
0 & 0 & U_1
\end{pmatrix}
,
H_{\rm c} =
\begin{pmatrix}
t & 0 & t \\ 
t & t & 0 \\
0 & t & t
\end{pmatrix}
.
\end{align}
The next key step is to incorporate the effect of $H_{\rm p}$, $H_{\rm c}$, and energy $E$ into the parameters of the effective model
\begin{align}
H_{\rm eff}' \psi_{\rm q} = E \psi_{\rm q},
\end{align}
where
\begin{align}
H_{\rm eff}' & = H_\text{q} + H_\text{c} {1\over E - H_\text{p}} H_\text{c}^\dagger \nonumber \\
& = \begin{pmatrix}
V_1 & 0 & 0 \\ 
0 & V_2 & 0 \\
0 & 0 & V_3
\end{pmatrix}
+
\begin{pmatrix}
t & 0 & t \\ 
t & t & 0 \\
0 & t & t
\end{pmatrix} 
\begin{pmatrix}
1/(E-U_1) & 0 & 0 \\ 
0 & 1/(E-U_1) & 0 \\
0 & 0 & 1/(E-U_1)
\end{pmatrix}
\begin{pmatrix}
t & t & 0 \\ 
0 & t & t \\
t & 0 & t
\end{pmatrix} \nonumber \\
& = \begin{pmatrix}
V_1+2t^2/(E-U_1) & t^2/(E-U_1) & t^2/(E-U_1) \\ 
t^2/(E-U_1) & V_2+2t^2/(E-U_1) & t^2/(E-U_1) \\
t^2/(E-U_1) & t^2/(E-U_1) & V_3+2t^2/(E-U_1)
\end{pmatrix}
.
\end{align}
Further simplification yields the eigenvalue equation of the effective model 
\begin{align}
\begin{pmatrix}
(E-U_1)V_1 & t^2 & t^2 \\ 
t^2 & (E-U_1)V_2 & t^2 \\
t^2 & t^2 & (E-U_1)V_3
\end{pmatrix}
\psi_{\rm q}
=
[(E-U_1)E-2t^2] \psi_{\rm q}
,
\end{align}
from which we identify the characteristic functions 
\begin{align}
g(E) = E-U_1, \quad f(E) = (E-U_1)E-2t^2.
\label{eq-fg}
\end{align}

Additionally, we can re-express the Hamiltonian in an alternative basis $(\psi_{{\rm q},1}, \psi_{{\rm p},1}, \psi_{{\rm q},2}, \psi_{{\rm p},2}, \psi_{{\rm q},3}, \psi_{{\rm p},3})^{\rm T}$
\begin{align}
H =
\begin{pmatrix}
V_1 & t & 0 & 0 & 0 & t \\ 
t & U_1 & t & 0 & 0 & 0 \\
0 & t & V_2 & t & 0 & 0 \\
0 & 0 & t & U_1 & t & 0 \\
0 & 0 & 0 & t & V_3 & t \\
t & 0 & 0 & 0 & t & U_1
\end{pmatrix}
\label{eq-hami2},
\end{align}
The corresponding eigenvalue equations give
\begin{align}
V_i \psi_{{\rm q},i} + t(\psi_{{\rm p},i-1} + \psi_{{\rm p},i}) = E \psi_{{\rm q},i}, \quad U_1 \psi_{{\rm p},i} + t(\psi_{{\rm q},i} + \psi_{{\rm q},i+1}) = E \psi_{{\rm p},i}
,
\label{eq-eigenvalue}
\end{align}
which leads to
\begin{align}
\psi_{{\rm p},i} = \frac{t(\psi_{{\rm q},i} + \psi_{{\rm q},i+1})}{E-U_1}
.
\end{align}
Substitute it into Eq. (\ref{eq-eigenvalue}), we can obtain
\begin{align}
V_i \psi_{{\rm q},i} + t(\frac{t(\psi_{{\rm q},i-1} + \psi_{{\rm q},i})}{E-U_1} + \frac{t(\psi_{{\rm q},i} + \psi_{{\rm q},i+1})}{E-U_1}) & = E \psi_{{\rm q},i},
\end{align}
which is reduced to 
\begin{align}
(E-U_1) V_i \psi_{{\rm q},i} + t^2(\psi_{{\rm q},i-1} + \psi_{{\rm q},i+1}) & = [(E-U_1) E - 2t^2] \psi_{{\rm q},i}.
\end{align}
The above result recovers the identical result in Eq. (\ref{eq-fg}). Thus the above two methods with different arrangement of bases, should lead to the same effective Hamiltonian.

\begin{figure*}
\includegraphics[width=0.95\textwidth]{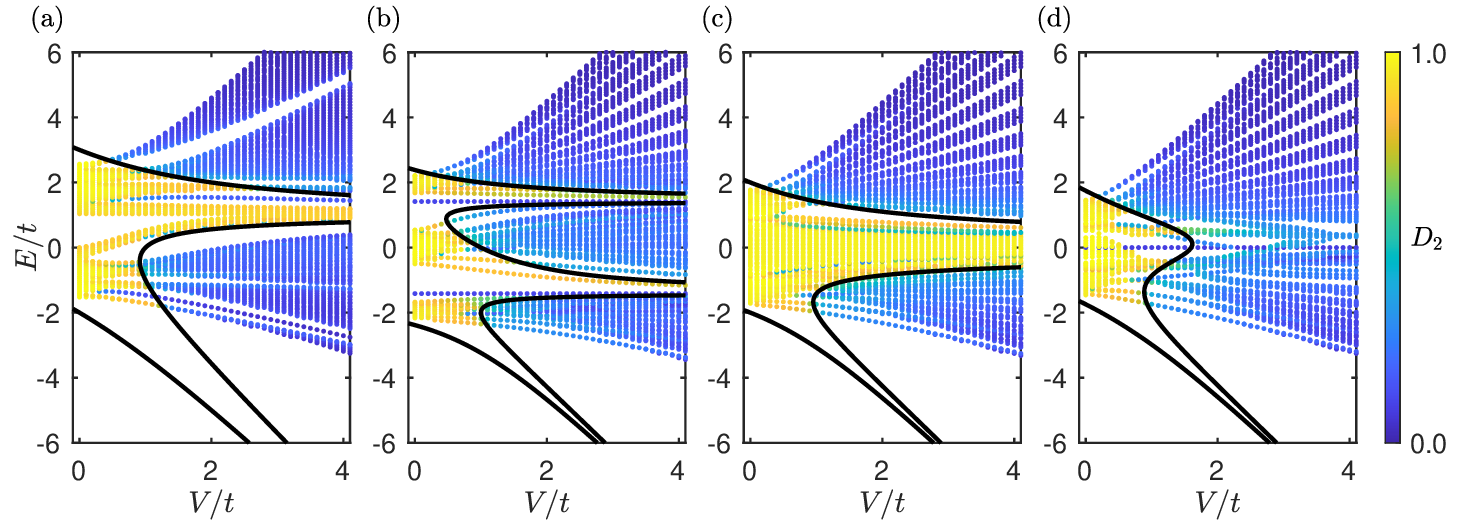}
\caption{\label{fig-figsup2} Verification of the generalized self-dual points with quasiperiodic mosaic models (a) $\kappa = 2$, $U_1 = t$, (b) $\kappa = 3$, $U_1 = -U_2 = t$, (c) $U_1 = -U_2 = it$, and (d) $U_1 = -U_2 = 1.5it$. The models in (c) and (d) have PT symmetry. The black lines in each figure denote the MEs given by Eq. (\ref{eq-S6}). The other parameters are $\alpha = (\sqrt{5} - 1)/2$, and $L = 987$.}
\end{figure*}

\section{Generalized self-dual points in two quasiperiodic potentials}
\label{sec-ApendixC}
In this section, we present a generalized self-duality framework for energy-dependent quasiperiodic systems, extending the results in the AAH potential.  Consider the following model with energy-dependent quasiperiodic potential (set $J = t^\kappa$)
\begin{align}
J (c_{i-1} + c_{i+1}) + g(E) V_i c_i = f(E) c_i
,
\label{eq-S1}
\end{align}
where the quasiperiodic potential $V_i = V \dfrac{{\rm cos}(2 \pi \alpha i + \phi)}{1 - b {\rm cos}(2 \pi \alpha i + \phi)}$ contains both phase $\phi$ and deformation parameter $b$ \cite{Ganeshan2015Mobility, Zhang2022Lyapunov, Wang2023mobility, Padhan2024Complete}. Introducing $\cosh \beta = \dfrac{1}{b}$, we can rewrite it as
\begin{align}
J (c_{i-1} + c_{i+1}) + h \chi_i(\beta) c_i = [f(E) + g(E) V \cosh \beta] c_i
,
\label{eq-S2}
\end{align}
where we have defined the function
\begin{align}
\chi_i(\beta) = \dfrac{\sinh \beta}{\cosh \beta - \cos(2\pi \alpha i + \phi)} = \sum_{r = - \infty}^{\infty} e^{-\beta |r|} e^{ir(2 \pi \alpha i + \phi)}
,
\label{eq-S3}
\end{align}
with $h = g(E) V \dfrac{\cosh \beta}{\tanh \beta}$. Under the dual transformation
\begin{align}
b_k = \sum_{mni} e^{i 2 \pi \alpha (km + mn + ni)} \chi_n^{-1} (\beta_0) c_i
,
\label{eq-S4}
\end{align}
where $2J \cosh \beta_0 = f(E) + g(E) V \cosh \beta$, we can map the original model to a dual equation
\begin{align}
J (b_{k-1} + b_{k+1}) + h \chi_k(\beta_0) b_k = 2J \cosh \beta b_k
.
\label{eq-S5}
\end{align}
Self-duality emerges when $\beta = \beta_0$, yielding the critical condition
\begin{align}
b f(E) = \pm 2 J - V g(E),
\label{eq-S6}
\end{align}
which has been used in the main text. When $f(E) = E$ and $g(E) = 1$, it is reduced to the well-known results in the literature \cite{Ganeshan2015Mobility}. 

\begin{figure*}
\includegraphics[width=0.95\textwidth]{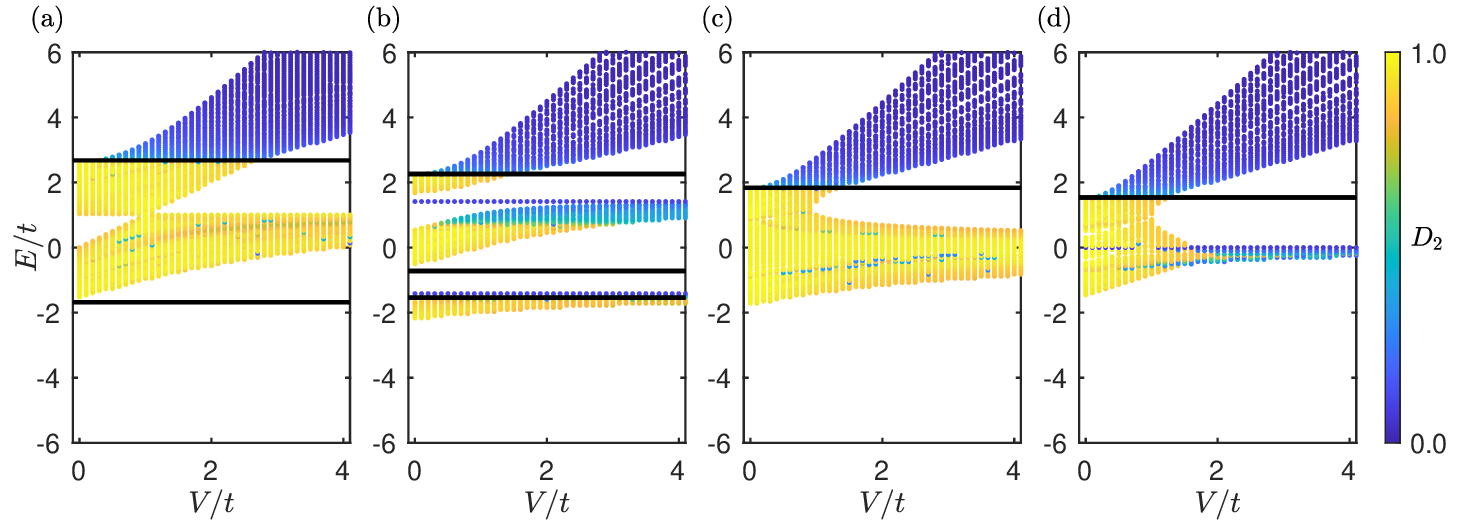}
\caption{\label{fig-figsup3} 
Verification of the generalized self-dual points with quasiperiodic mosaic models (a) $\kappa = 2$, $U_1 = t$, (b) $\kappa = 3$, $U_1 = -U_2 = t$, (c) $U_1 = -U_2 = it$, and (d) $U_1 = -U_2 = 1.5it$. The models in (c) and (d) have PT symmetry. The black lines in each figure denote the MEs given by Eq. (\ref{eq-S10}). The other parameters are $\alpha = (\sqrt{5} - 1) / 2$, and $L = 987$.}
\end{figure*}

For the quasiperiodic non-Hermitian potential with $V_i = \dfrac{V}{1 - b e^{i 2 \pi \alpha i}}$ \cite{Liu2020Generalized, Liu2022transition}, we can also rewrite Eq. (\ref{eq-S1}) as
\begin{align}
2 \sinh{(s)} \Omega_m (s) \phi_m = g(E) V \sum_{r'} e^{-|m-r'|s} \phi_{r'}
,
\label{eq-S7}
\end{align}
with 
\begin{align}
\phi_m = \sum_i e^{i 2 \pi \alpha m i} c_i, \hspace{0.3cm} 2 \cosh{(s)} = f(E), \hspace{0.3cm} \Omega_m(s) = \dfrac{\cosh(s) - \cos(2 \pi \alpha m)}{\sinh(s)}
.
\label{eq-S8}
\end{align} With the dual transformation $\phi_k = \sum_{m} e^{i 2 \pi \alpha mk} \Omega_m(s) \phi_m$, we obtain
\begin{align}
2 \sinh{(s)} \Omega_k (s_0) \phi_k = g(E) V \sum_{k'} e^{-|k-k'|s} \phi_{k'}
,
\label{eq-S9}
\end{align}
where $e^{s_0} = \dfrac{1}{b}$. This self-dual condition is determined by $s = s_0$, yielding
\begin{align}
f(E) = b + \dfrac{1}{b}
.
\label{eq-S10}
\end{align}
To verify the generalized self-dual points given by Eqs. (\ref{eq-S6}) and (\ref{eq-S10}), we present the fractal dimension of the eigenstates as a function of the potential strength for various quasiperiodic mosaic models in Figs. \ref{fig-figsup2} and \ref{fig-figsup3}. The black lines in the figures represent the MEs obtained from the generalized self-dual points. The other parameters are $\alpha = (\sqrt{5} - 1) / 2$, and $L = 987$. These results confirm the theoretical predictions.

\section{Two steps to obtain the MEs in quasiperiodic network models}
\label{sec-ApendixD}
In this section, we present the details for deriving the MEs in various quasiperiodic network models based on hidden self-duality. This process is carried out in two steps.  
\begin{enumerate}
    \item \textbf{Effective Hamiltonian construction via  integrating out periodic sites} 

    For the simplified model in Fig.~\ref{fig-figsup1}(a) [the same procedure for Fig.~\ref{fig-figsup1}(b)], we can treat the incommensurate sites as boundary condition, and formulate the equation of motion for the periodic sites ($d_{i,k}$) as  
    \begin{equation}
    U_{k} d_{i, k} + \sum_{k'} t_{k,k'} d_{i, k'} = A_{k,i} c_i +  B_{k, i+1} c_{i+1}.
    \end{equation}
    In this equation, $U_{k}$ is the potential of the $k$-th periodic site, $t_{k,k'}$ is the hopping term between different periodic sites, and $A_{k,i}$ and $B_{k,i+1}$ couple periodic sites ($d_{i,k}$) to adjacent incommensurate sites ($c_i$, $c_{i+1}$). Applying the boundary condition projection method detailed in Sec.~\ref{sec-ApendixB}, we eliminate $c_i$ variables and derive an effective quasiperiodic chain for $c_i$ sites 
    \begin{align}
    J (c_{i-1} + c_{i+1}) + g(E) V_i c_i = f(E) c_i.
    \end{align}
    Here, energy-dependent function $g(E)$ and $f(E)$ emerge as rescaling factors: $f(E)$ rescales the effective energy while $g(E)$ modulates the quasiperiodic potential strength $V_i$. This reduction procedure has been systematically implemented for three distinct network geometries, with corresponding $f(E)$ and $g(E)$ functions cataloged in Table I of the main text.

    \item \textbf{Identifying the self-dual condition}

    The derived effective Hamiltonian permits universal MEs determination through self-duality analysis. If we choose a incommensurate potential $V_j$ with generalized self-dual points at 
    \begin{equation}
    F(f(E), g(E)V) = 0,
    \end{equation}
then what we need to do is to substitute the specific functions $f(E)$ and $g(E)$ of the effective Hamiltonian into the above self-dual condition to obtain the concrete MEs. In this work, we have considered the following three incommensurate potentials 
    \begin{eqnarray} 
V_i &=& V {\rm cos}(2\pi \alpha i), \\ 
V_i &=& {V \cos(2\pi \alpha i) \over 1 - b \cos(2\pi \alpha i)}, \\ 
V_i &=& {V \over 1 - b \exp(i 2\pi \alpha i)},
\end{eqnarray}
which respectively yield MEs at
\begin{eqnarray}
g(E) V &=& \pm 2J,  \\ 
bf(E) &=& \pm 2J - V g(E), \\ 
f(E) &=& b + {1 \over b}.
\end{eqnarray}
\end{enumerate}

\subsection{Generalization to long-range hopping systems} 

Although the above analysis mainly focus on models with nearest-neighbor hopping, this method in principle works equally well for solvable models with long-range hopping. Consider network models that can be mapped to the following effective quasiperiodic chain \cite{biddle2010mobility}
\begin{align}
\sum_{i\neq i'} t e^{-p|i-i'|} c_{i'} + g(E)V\cos(2\pi\alpha i) c_i = f(E)c_i
,
\end{align}
where $p$ governs the hopping range. The hidden self-duality ensures the MEs at
\begin{align}
\cosh(p) = \frac{f(E)+t}{g(E)V},
\end{align}
This nonlinear equation can yield rich roots by varying of $E$ and $p$.

\section{Verification of the MEs in two quasiperiodic network models discussed in the main text}
\label{sec-ApendixE}
In the section, we validate our general method through the other two quasiperiodic network models from the main text.

\subsection{Side-coupled quasiperiodic chain}
\label{sec-ApendixE-1}

We begin with the side-coupled quasiperiodic chain, described by the Hamiltonian
\begin{align}
    H = \sum_i (V_i c_i^\dagger c_i + \sum_{k=1}^2 U_k d_{i,k}^\dagger d_{i,k}) - t \sum_i (c_{i+1}^\dagger c_i + \sum_{k=1}^2 c_i^\dagger d_{i,k} + {\rm H.c.}),
\end{align}
corresponding to the second structure in Table I.

\subsubsection{Rescaling function}
The motion equations of periodic sites are
\begin{align}
U_1 d_{i,1} -tc_{i} = E d_{i,1}, \quad U_2 d_{i,2} - tc_{i} = E d_{i,2},
\end{align}
which yield
\begin{align}
d_{i,1} = \frac{t}{U_1 - E} c_{i}, \quad
d_{i,2} = \frac{t}{U_2 - E} c_{i}.
\end{align}
Substitute them into the the equation of $c_i$
\begin{align}
V_i c_{i} -t (d_{i,1}+d_{i,1}+c_{i-1}+c_{i+1}) = E c_{i},
\end{align}
we can obtain the following effective equation
\begin{align}
V_i c_i -t (c_{i-1}+c_{i+1}) = (E-\frac{t^2}{E-U_1}-\frac{t^2}{E-U_2}) c_i
.
\end{align}
This indicates the rescaling function
\begin{align}
f(E) = E-t^2/(E-U_1)-t^2/(E-U_2), \quad g(E) = 1.
\end{align}

\subsubsection{ME analysis for three quasiperiodic classes}
For AAH potential $V_i = V \cos(2 \pi \alpha i)$, the generalized self-dual points at $g(E)V= \pm2t$ yield the same extended-localized transition point $V = \pm 2t$ as in the classical AAH chain, indicating the absence of MEs in this side-coupled quasiperiodic AAH model.

In contrast, for a quasiperiodic potential $V_i = V\cos(2\pi\alpha i)/[1-b\cos(2\pi\alpha i)]$, the self-dual points $bf(E) = \pm 2t-Vg(E)$ lead to the MEs at
\begin{align}
    E = \frac{2-V\pm\sqrt{4+8b^2-4V+V^2}}{2b}, \quad E = \frac{-2-V\pm\sqrt{4+8b^2+4V+V^2}}{2b},
\end{align}
where we set $U_1 = U_2 = 0$ and $t = 1$. This result indicates the existence of four MEs across the whole $V$ region. For nonzero $U_1$ and $U_2$, the method is similar. 

For potential given by $V_i = V/[1-be^{i2\pi\alpha i}]$, the self-dual points $f(E) = b+1/b$ yield the MEs at
\begin{align}
    E = \frac{1+b^2\pm\sqrt{1+10b^2+b^4}}{2b},
\end{align}
revealing two MEs that are independent of the potential strength.

\begin{figure*}
\includegraphics[width=0.95\textwidth]{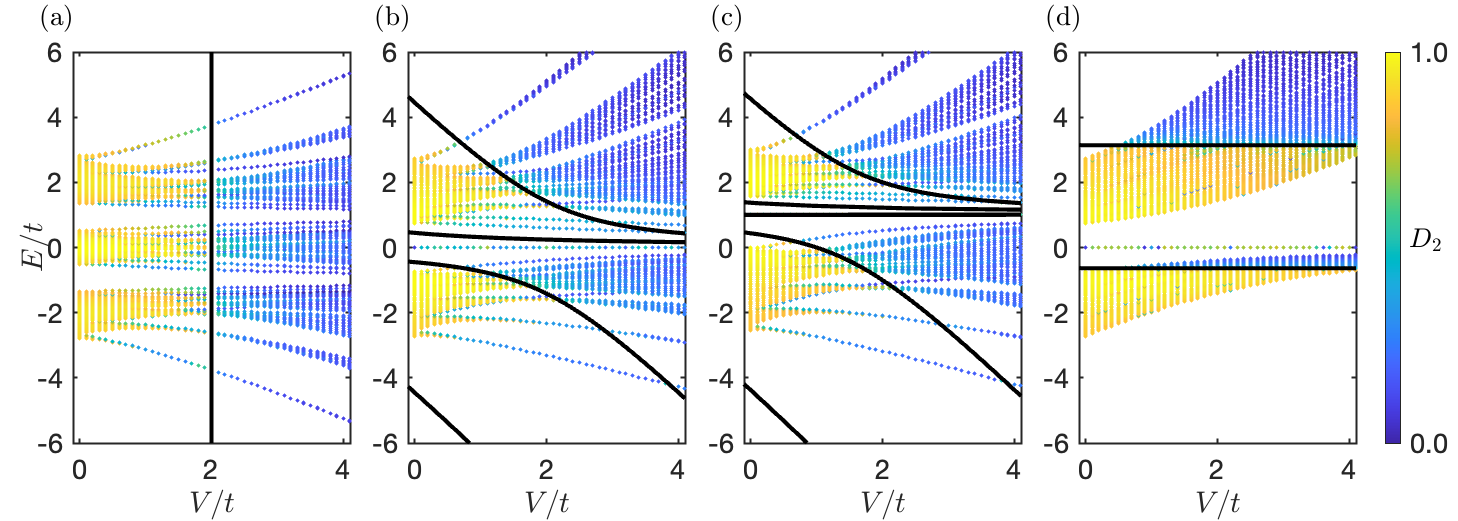}
\caption{\label{fig-figsup4} 
The MEs in side-coupled quasiperiodic chain with various quasiperiodic potentials for (a) $U_1 = U_2 = 0$, $V_i = V\cos(2\pi\alpha i)$, (b) $U_1 = U_2 = 0$, $V_i = V\cos(2\pi\alpha i)/[1-b\cos(2\pi\alpha i)]$, (c) $U_1 = U_2 = t$, $V_i = V\cos(2\pi\alpha i)/[1-b\cos(2\pi\alpha i)]$, and (d) $U_1 = U_2 = 0$, $V_i = V/[1-e^{i2\pi\alpha i}]$. Here the black lines denote the MEs obtained from the  self-dual points. The other parameters are $\alpha = (\sqrt{5} - 1) / 2$, and $L = 600$.}
\end{figure*}

To verify these results, we plot the fractal dimension of eigenstates versus potential strength for the side-coupled quasiperiodic chain with various quasiperiodic potentials in Fig. \ref{fig-figsup4}. We set $U_1 = U_2 = 0$ in Figs. \ref{fig-figsup4}(a), \ref{fig-figsup4}(b), and \ref{fig-figsup4}(d), and $U_1 = U_2 = t$ in Fig. \ref{fig-figsup4}(c). The MEs agree well with the numerical results.

\begin{figure*}
\includegraphics[width=0.95\textwidth]{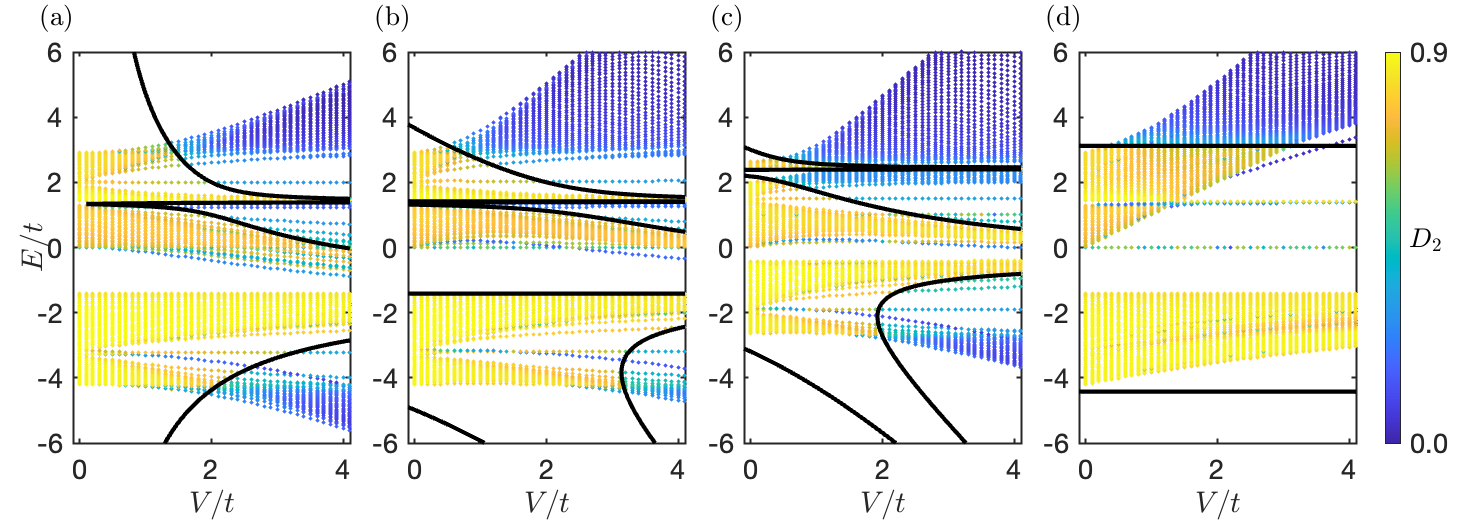}
\caption{\label{fig-figsup5} 
The MEs in multipath quasiperiodic chain with various quasiperiodic potentials for (a) $U = 0$, $t_1 = t$, $V_i = V\cos(2\pi\alpha i)$, (b) $U= 0$, $t_1 = t$, $V_i = V\cos(2\pi\alpha i)/[1-b\cos(2\pi\alpha i)]$, (c) $U = t$, $t_1 = 0.5t$, $V_i = V\cos(2\pi\alpha i)/[1-b\cos(2\pi\alpha i)]$, and (d) $U = 0$, $t_1 = t$, $V_i = V/[1-e^{i2\pi\alpha i}]$. Here the black lines denote the MEs obtained from the  self-dual points. The other parameters are $\alpha = (\sqrt{5} - 1) / 2$, and $L = 600$.}
\end{figure*}

\subsection{Multipath quasiperiodic chain}
\label{sec-ApendixE-2}
The third configuration in Table I features a multipath structure with Hamiltonian
\begin{align}
    H = & \sum_i (V_i c_i^\dagger c_i + \sum_{k=1}^3 U_k d_{i,k}^\dagger d_{i,k}) - t \sum_i (c_{i}^\dagger d_{i,2} + c_{i+1}^\dagger d_{i,2} + \sum_{k=1}^2 d_{i,k}^\dagger d_{i,k+1} + {\rm H.c.}) \nonumber \\
    & - t_1 \sum_i \sum_{k = 1, 3} (c_{i}^\dagger d_{i,k} + c_{i+1}^\dagger d_{i,k} + {\rm H.c.}).
\end{align}
For simplification, we set $U_k = U$ here.

\subsubsection{Rescaling function}
The motion equation for periodic sites take the form 
\begin{align}
Ud_{i-1,1} -t_1(c_{i-1}+c_{i}) - td_{i-1,2} & = Ed_{i-1,1}, \nonumber \\
Ud_{i-1,2} - t(d_{i-1,1}+d_{i-1,3}+c_{i-1}+c_{i}) & = Ed_{i-1,2}, \nonumber \\
Ud_{i-1,3} -t_1(c_{i-1}+c_{i}) - td_{i-1,2} & = Ed_{i-1,3}, \nonumber \\
Ud_{i,1} -t_1(c_{i}+c_{i+1}) - td_{i,2} & = Ed_{i,1} \nonumber, \\
Ud_{i,2} - t(d_{i,1}+d_{i,3}+c_{i}+c_{i+1}) & = Ed_{i,2}, \nonumber \\
Ud_{i,3} -t_1(c_{i}+c_{i+1}) - td_{i,2} & = Ed_{i,3}.
\end{align}
From these equations, we derive
\begin{align}
d_{i-1,1} & = \frac{(Et_1-Ut_1-t^2)c_{i-1}+(Et_1-Ut_1-t^2)c_{i}}{2t^2-E^2+2EU-U^2}, \nonumber \\
d_{i-1,2} & = \frac{(E-2t_1-U)c_{i-1}+(E-2t_1-U)c_{i}}{2t^2-E^2+2EU-U^2}, \nonumber \\
d_{i-1,3} & = \frac{(Et_1-Ut_1-t^2)c_{i-1}+(Et_1-Ut_1-t^2)c_{i}}{2t^2-E^2+2EU-U^2}, \nonumber \\
d_{i,1} & = \frac{(Et_1-Ut_1-t^2)c_{i}+(Et_1-Ut_1-t^2)c_{i+1}}{2t^2-E^2+2EU-U^2}, \nonumber \\
d_{i,2} & = \frac{(E-2t_1-U)c_{i}+(E-2t_1-U)c_{i+1}}{2t^2-E^2+2EU-U^2}, \nonumber \\
d_{i,3} & = \frac{(Et_1-Ut_1-t^2)c_{i}+(Et_1-Ut_1-t^2)c_{i+1}}{2t^2-E^2+2EU-U^2}.
\end{align}
Substitute the expressions of $d_{i,k}$ into the equation for $c_i$
\begin{align}
V_ic_{i} -t_1(d_{i-1,1}+d_{i-1,3}+d_{i,1}+d_{i,3}) - t(d_{i-1,2}+d_{i,2}) = Ec_{i},
\end{align}
we have the following effective equation
\begin{align}
    g(E)V_i c_i + J (c_{i-1}+c_{i+1}) = f(E) c_i,
\end{align}
with rescaling function
\begin{eqnarray}
    g(E) &=& \frac{E^2-2UE+U^2-2t^2}{(2t_1^2+t^2)E-Ut^2-4t_1t^2-2t_1^2U}, \\
    f(E) &=& \frac{E^3-2UE^2+(U^2-4t_1^2-4t^2)E+2Ut^2+4t_1(t^2+t_1U)}{(2t_1^2+t^2)E-Ut^2-4t_1t^2-2t_1^2U}.
\end{eqnarray}

\subsubsection{ME analysis for three quasiperiodic classes}
For AAH potential $V_i = V \cos(2 \pi \alpha i)$, substitute $g(E)$ into the self-dual points $g(E)V=\pm2J$ leads to the MEs at
\begin{align}
E = \frac{3\pm\sqrt{9-8V+2V^2}}{V}, \quad E = \frac{-3\pm\sqrt{9+8V+2V^2}}{V},
\end{align}
where we set $U$ $= 0$ and $t_1 = t$ for simplicity. Note that the denominator of the above equations diverges when $V$ tends to be $0$. Applying L'Hôpital's rule, the MEs degenerate to $E = \infty$, $-\infty$, and $4t/3$ when $V \rightarrow 0$. 

Similarly, for $V_i = V \cos(2\pi\alpha i)/[1-b\cos(2\pi\alpha i)]$, the self-dual conditions $bf(E) = \pm2J-Vg(E)$ give the MEs at
\begin{align}
    V = \frac{-bE^3+8bE-8b\pm(8-6E)}{E^2-2}.
\end{align}
Here, we observe two special energies $E = \pm \sqrt{2}t$, for which the critical potential strength are infinite. 

Finally, for $V_i = V/[1-be^{i2\pi\alpha i}]$, self-dual points $f(E) = b +1/b$ yields the ME at
\begin{align}
\frac{E^3-8E+8}{3E-4} = b + \frac{1}{b},
\end{align}
which is independent of the potential.

In Fig.~\ref{fig-figsup5}, we present some numerical evidence supporting the above conclusions. We set $U = 0$, $t_1 = t$ in Figs.~\ref{fig-figsup5}(a), \ref{fig-figsup5}(b), and \ref{fig-figsup5}(d), and $U = t$, $t_1 = 0.5t$ in Fig.~\ref{fig-figsup5}(c). The numerical results agree well with the analytical MEs.

\begin{figure}[tbp]
	\centering
	\includegraphics[width=0.95\textwidth]{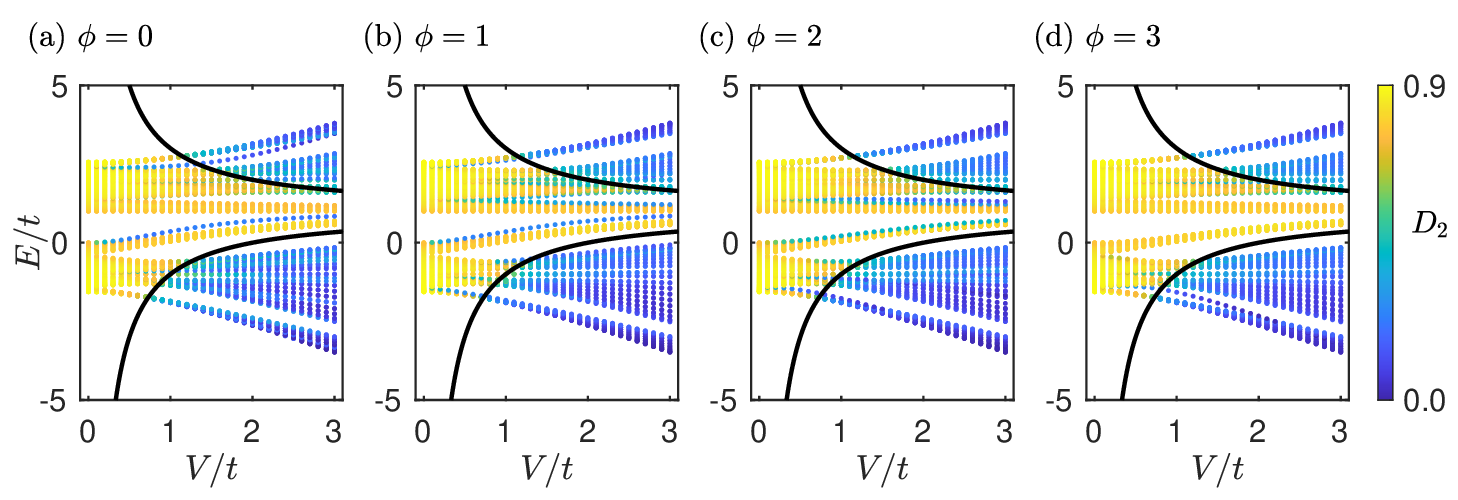}
	\caption{Fractal dimension $D_2$ as a function of energy $E$ and potential strength $V$ when (a) $\phi = 0$, (b) $\phi = 1$, (c) $\phi = 2$, and (d) $\phi = 3$. Here we choose quasiperiodic mosaic models with $\kappa = 2$ and $U_1 = t$ as an example, where the MEs are given by $E=\pm2t/V+t$.}
	\label{fig-figsup6}
\end{figure}
\begin{figure}[tbp]
	\centering
	\includegraphics[width=0.5\textwidth]{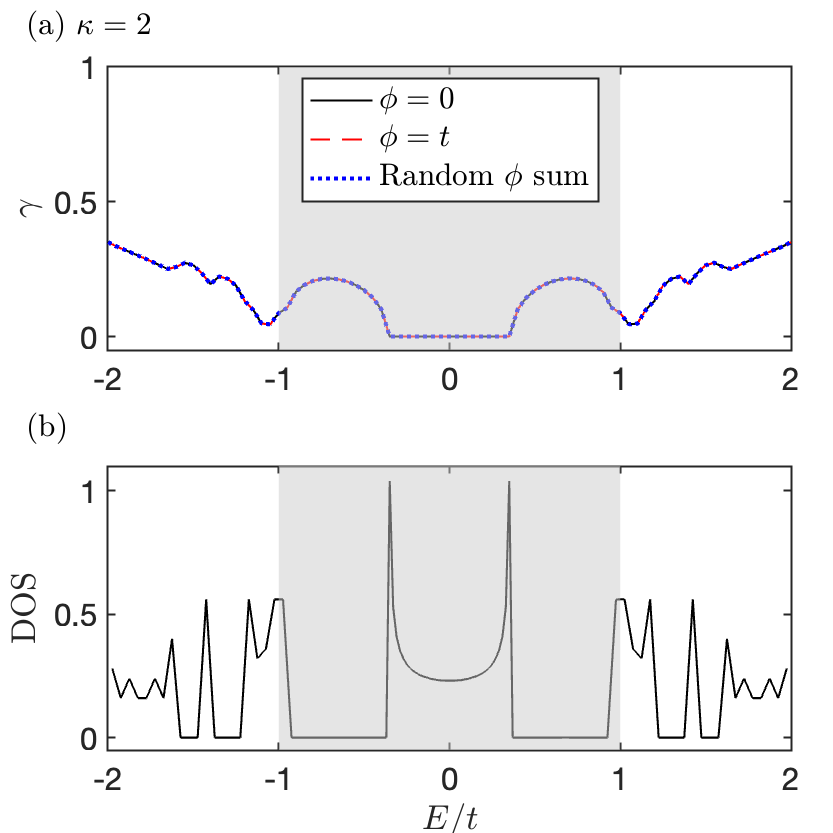}
	\caption{(a) Lyapunov exponent $\gamma$ versus energy $E$ under different phase term. (b) The corresponding DOS versus $E$. The gray area denotes the extended phase $|E|<t$.}
	\label{fig-figsup7}
\end{figure}

\section{Phase independence of MEs}
\label{sec-ApendixF}
In this section, we investigate the effect of the phase term $\phi$ on the localization properties through analysis of fractal dimension ($D_2$), Lyapunov exponent ($\gamma$), and density of states (DOS). Our results confirm that the phase term doesn't influence the localization-delocalization transition points, in agreement with previous results. Specifically, we present the fractal dimension $D_2$ versus energy $E$ and potential strength $V$ under different values of $\phi$ in Fig.~\ref{fig-figsup6}. It is observed that the MEs are located at the same position, although the eigenenergy spectrum are changed by $\phi$. Here we consider the quasiperiodic mosaic models with $\kappa = 2$ and $U_1 = t$ as an example, with the MEs located at $E = \pm 2t/V+t$. In addition, we plot the Lyapunov exponent $\gamma$ and DOS as functions of energy $E$ in Fig.~\ref{fig-figsup7}. In Fig.~\ref{fig-figsup7}(a), the blue dotted line represents the results averaged over 100 realizations. Our findings indicate that the Lyapunov exponent remains unaffected by $\phi$, indicating that the localization properties are independent of the phase term $\phi$. Additionally, for energies satisfying $|E|<t$ and $D(E) \neq 0$, we observe Lyapunov exponent $\gamma = 0$, consistent with our expectation. Here we choose the mosaic model with $\kappa = 2$ and $U_1 = 0$ as an example, where the MEs should be located at $E=\pm t$. 

 \begin{figure}[tbp]
	\centering	\includegraphics[width=0.8\textwidth]{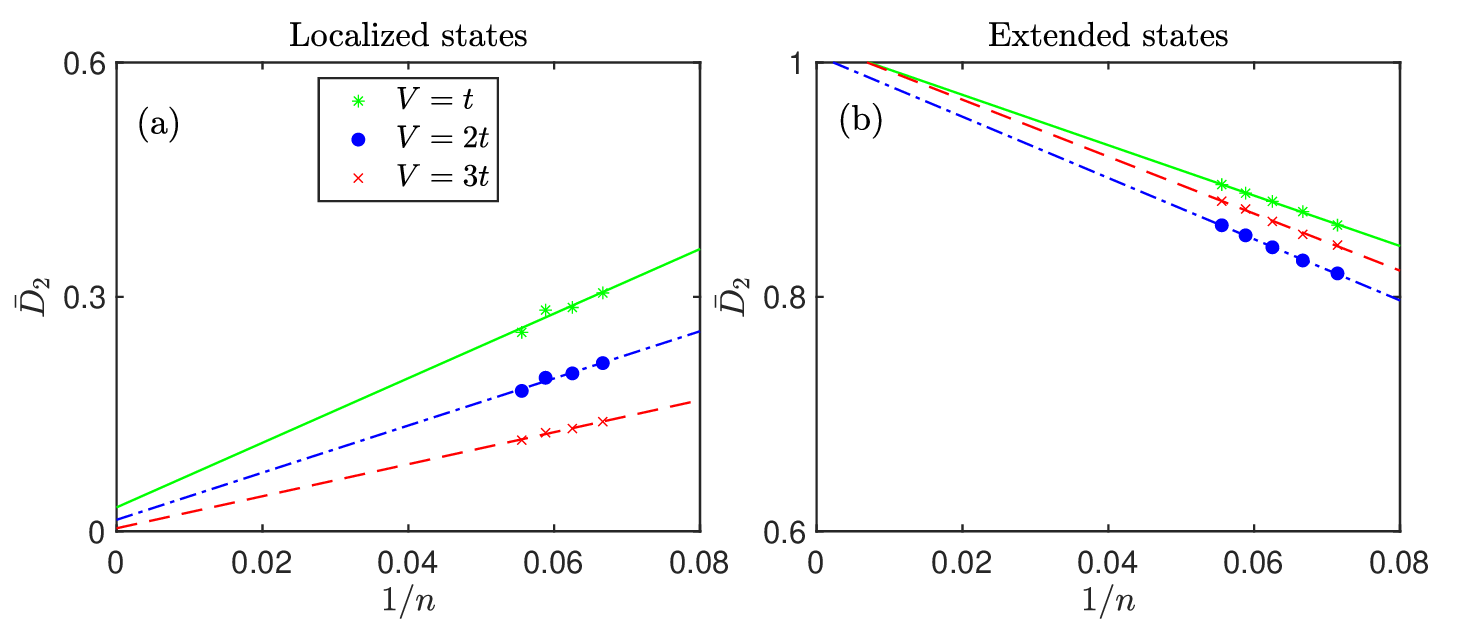}
	\caption{Mean fractal dimension $\bar{D}_2$ versus $1/n$ for $(a)$ localized and $(b)$ extended states under different potential strength, where $n$ denotes the Fibonacci numbers. Here we choose quasiperiodic mosaic models with $\kappa = 2$ and $U_1 = t$ as an example, where the extended states satisfy $-2t/V+U_1<E<2t/V+U_1$}
	\label{fig-figsup8}
\end{figure}

\section{Finite size scaling}
\label{sec-ApendixG}
In this section, we investigate the finite-size scaling. As a representative case, we consider the quasiperiodic mosaic models with $\kappa = 2$ and $U_1 = t$, where the extended states satisfying $-2t/V+U_1<E<2t/V+U_1$. In Fig.~\ref{fig-figsup8}, we present the mean fractal dimension $\bar{D}_2$ versus $1/n$ for (a) localized (b) extended states under different potential strengths. In our numerical calculations, the chain length is chosen as the Fibonacci number $F_n$, with the incommensurate modulation parameter set to $\alpha = F_{n-1}/F_n$, where $F_n$ is defined by $F_{n+1} = F_n + F_{n-1}$ with $F_0 = F_1 = 1$. As expected, $\bar{D}_2$ approaches 0 for localized states and 1 for extended states in the thermodynamic limit.

\bibliography{ref.bib}

\end{document}